\documentstyle[11pt,paspconf,epsf]{article}

\begin{document}

\title{OH(1720 MHz) Masers at the Galactic Centre}

\author{Mark Wardle}
\affil{Special Research Centre for Theoretical Astrophysics, 
University of Sydney, NSW 2006, Australia}

\author{Farhad Yusef-Zadeh}
\affil{Department of Physics \& Astronomy, Northwestern University, 
2131 N. Sheridan Rd., Evanston, IL 60208, USA}

\author{T.R. Geballe}
\affil{Joint Astronomy Centre, 660 N. A'ohoku Pl., Hilo, HI 96720, USA}

\begin{abstract}
OH(1720 MHz) masers permit direct measurements of the magnetic field 
strength at the Galactic Centre, and their angular broadening 
constrains models for the medium that scatters radio waves in the 
central 100 pc of the Galaxy.  As the 1720 MHz masers are 
unaccompanied by the main-line masers at 1665 and 1667 MHz, they must 
be pumped by collisions in molecular gas that is cooling after being 
overrun by a non-dissociative (i.e.  C-type) shock wave.  In 
particular, this confirms that the ``SNR'' Sgr A East is driving a 
shock into the M-0.02-0.07 molecular cloud.  The intensity of the 
v=1--0 S(1) line of H$_2$ is consistent with the shock strength 
expected to be driven into the molecular gas by the pressure within 
Sgr A East.
\end{abstract}

\index{masers}
\index{magnetic fields}
\index{interstellar scattering}
\index{shock waves}
\index{Source!M-0.02-0.07}
\index{Source!Sgr A East}

\section{Introduction}

OH masers are ``point'' probes of the gas in molecular clouds and star 
forming regions, allowing precise velocity measurements and Zeeman 
determinations of the magnetic field strength.  At the Galactic 
Centre, angular broadening of maser spots is also of interest for 
understanding the location and nature of the ionised medium that 
scatters radio waves.  If the conditions for level inversion are 
understood then in principle temperatures and densities in the gas, 
and some information about the radiation field can be inferred.  It 
turns out that the inversion of the 1720 MHz transition of OH requires 
a restricted set of physical conditions that can only be achieved in 
molecular gas that is cooling after being overrun by a 
non-dissociative (C-type) shock.  Therefore the masers mark a subset 
of the locations at the Galactic Centre where molecular gas is being 
violently disturbed.

\begin{figure}[!t]
	\plotone{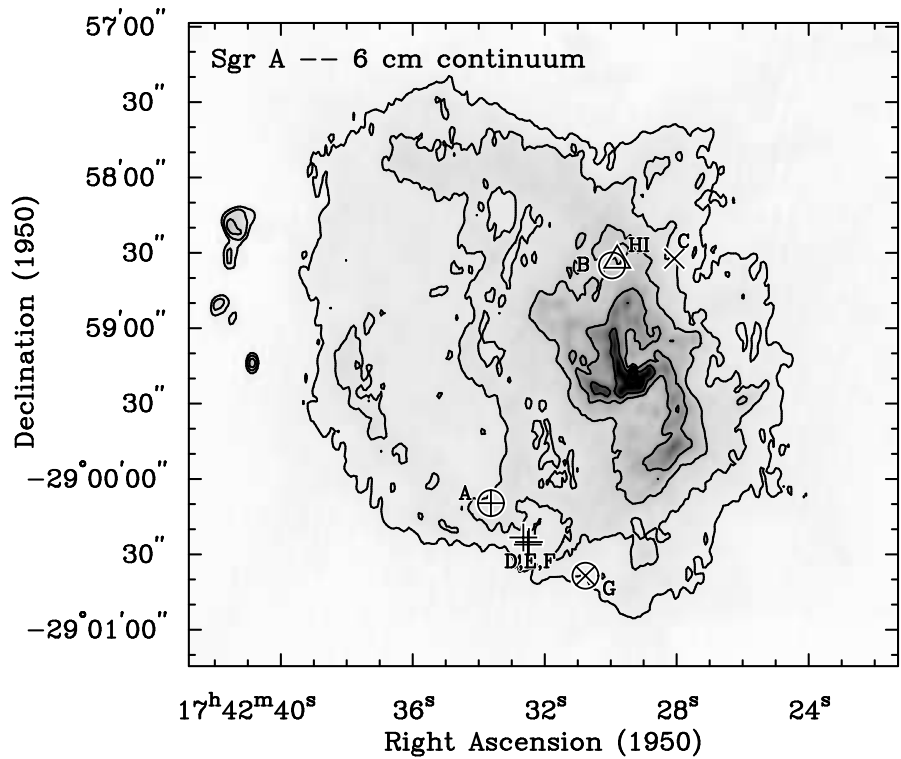} 
	\caption{OH(1720 MHz) 
	masers at the Galactic centre (Yusef-Zadeh et al. 1996).  The 
	maser positions are labelled A--G, and the location of an H\textsc{i} 
	Zeeman measurement by Plante et al. (1995) is labelled `H\textsc{i}'}
	\label{fig:YZ96masers}
\end{figure}

Several OH(1720 MHz) masers have been detected in Sgr A (Yusef-Zadeh 
et al. 1996); their locations are shown in Fig.  \ref{fig:YZ96masers}.
The gross dynamics of molecular gas associated with the Sgr A region 
have been studied in CO, HCN and CS (Serabyn et al. 1986; G\"usten et 
al. 1987; Serabyn, Lacy \& Achtermann 1992) so the maser velocities 
unambiguously associate them with the Galactic Centre.  The features A 
and D--G at velocities of +50--60 km/s are located within the 
M-0.02-0.07 (``50 km/s'') cloud, and B (+130 km/s) is located within 
the northern part of the circumnuclear ring.  Feature `C' at +43 km/s 
may also be physically associated with the northwest side of Sgr A 
East.

The masers allow an accurate determination of the magnetic field 
strength associated with the molecular gas at the Galactic Centre.  
Previous Zeeman absorption measurements, in H\textsc{i} (Schwarz et al. 1990; 
Plante, Lo \& Crutcher 1995) and in OH (Killeen, Lo \& Crutcher 1992), 
detected milliGauss field strengths, but the weak signal had to be 
combined over large regions.  These measurements are confounded by 
spatial gradients in line profile and probably by changes in the 
magnetic field along the line of sight (e.g.  through the 
circumnuclear ring, Wardle \& K\"onigl 1990).  The compactness of the 
masers avoids these problems, and Zeeman measurements by Yusef Zadeh et 
al. (1996) yield line-of-sight fields of $+$3mG in the M-0.02-0.07 
sources, $-$(3--4) mG for source B, and +2 mG for source C. In 
principle, the inferred line-of-sight field strength is overestimated 
if the masers are extremely saturated (Elitzur 1998), but this is not 
expected for OH(1720 MHz) masers in general (Lockett, Gauthier \& 
Elitzur 1998), and does not appear to be the case at the Galactic 
Centre (Yusef-Zadeh et al.  1998).

\section{Interstellar scattering}

Studies of radio-wave scattering at the Galactic Centre have 
previously been restricted to Sgr A$^{\mathrm{*}}$ (e.g.  Backer 1988; 
Yusef-Zadeh et al.  1994) and OH/IR stars (van Langevelde et al.  
1992; Frail et al.  1994).  These studies indicate that the scattering 
medium probably lies within 100 pc of the centre.  As the nearest 
OH/IR star is about 3 arcminutes away from Sgr A$^{\mathrm{*}}$, the 
OH(1720 MHz) masers usefully sample the properties of the screen along 
lines of sight that are closer in projection on the plane of the sky.  
The broadening of a source depends on proximity to the scattering 
medium, so it is helpful that the location of the masers along the 
line of sight is well-determined, unlike that of individual OH/IR 
stars which may lie 100 pc in front or behind Sgr A*.

The masers suffer an angular broadening similar to that of
Sgr A*, about an arcsecond at these frequencies (Yusef-Zadeh et al. 1998).
\begin{figure}[!t]
	\plotone{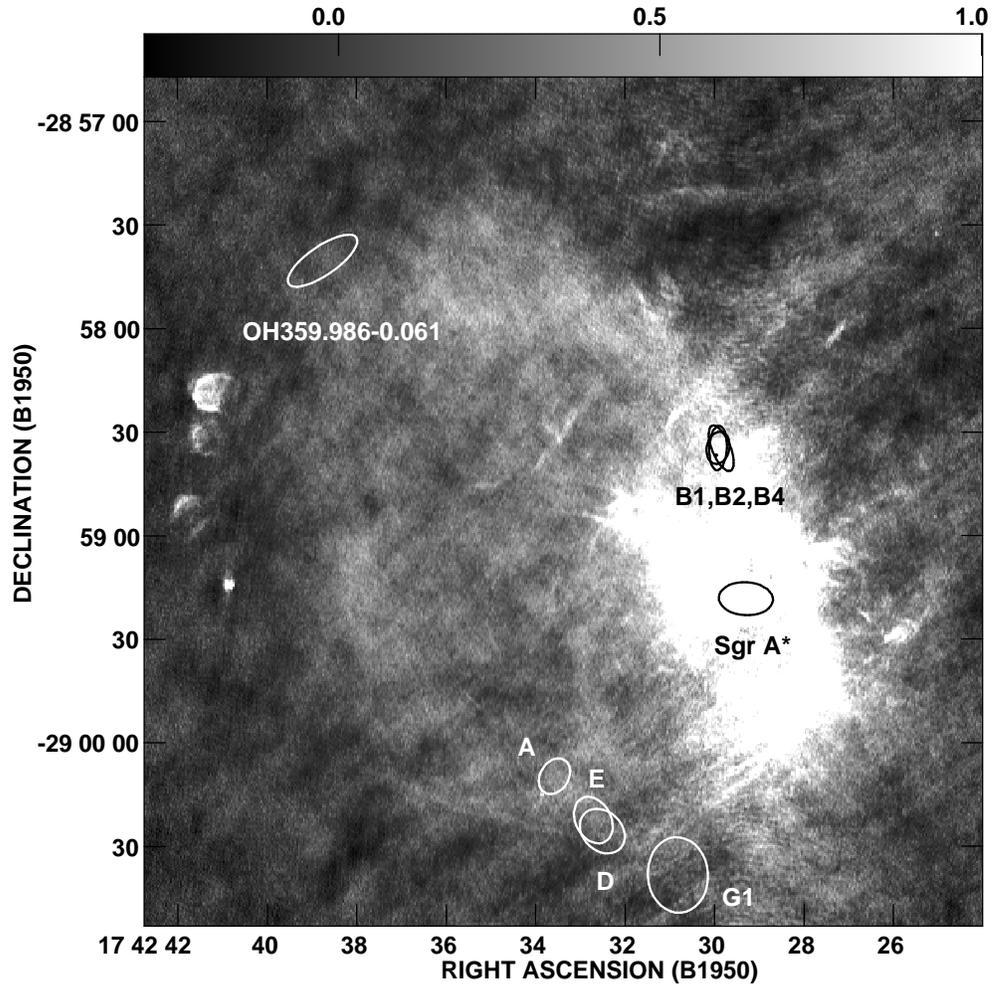} 
\caption{A $\lambda$6cm radiograph of Sgr A showing the scattering 
disks (enlarged by a factor of 15) of Sgr A$^{\mathrm{*}}$, an OH/IR 
star, and the OH(1720 MHz) masers (Yusef-Zadeh et al. 1998).}
	\label{fig:YZ98scattered}
\end{figure}
The scattering disks (see Fig.  \ref{fig:YZ98scattered}) differ in 
both ellipticity and position angle because of variations in the 
line-of-sight-integrated properties of the medium.  The conclusions 
that can be drawn are limited by the small number of samples 
available, but it is significant that the shape and orientation of the 
broadening differs between lines of sight separated by less than a 
parsec -- if the intervening scattering medium were thicker than a few 
parsecs the integration along the line of sight would tend to wash 
these differences out.  This favours ``interface'' models (Lazio \& 
Cordes 1998) over, for example, the $10^{8}\,$K gas that fills the 
inner half a degree of the Galaxy (Yamauchi et al.  1990).  Further, 
maser C does not show the same degree of broadening as the other 
sources, implying that if it is located at the Galactic Centre then 
the scattering screen must be patchy.

\section{OH(1720 MHz) masers and shock waves}

OH masers have traditionally been associated with H\textsc{ii} regions and 
evolved stars.  However, it has recently become apparent that 1720 MHz 
OH masers unaccompanied by 1665 and 1667 MHz masers are produced where 
supernova remnants (SNRs) interact with molecular clouds (e.g.  Frail, 
Goss \& Slysh 1994; Claussen et al.  1997; Koralesky et al.  1998).  
This is supported by studies of the pumping of the OH maser lines 
(Elitzur 1976; Pavlakis \& Kylafis 1996a,b; Lockett et al.  1998), 
which show that the OH(1720 MHz) maser is collisionally pumped in 
molecular gas at temperatures and densities between 50-125 K and 
$10^5-10^6$ cm$^{-3}$ respectively, whereas the 1665 and 1667 MHz 
masers are generally pumped by far-infrared photons.  Thus the 1665/7 
MHz masers are associated with H\textsc{ii} regions and evolved stars.  
Occasionally 1720 MHz masers are also associated with main-line 
masers, but in their absence are located in cooling, shocked molecular 
gas under specific physical conditions.

Although the pumping conditions for the 1720 MHz OH masers suggest 
that they are associated with shock waves, OH itself is not 
formed directly by shock chemistry (e.g.  Draine, Roberge \& Dalgarno 
1983; Hollenbach \& McKee 1989; Kaufman \& Neufeld 1996).  Instead, 
the OH must be produced by dissociation of the copious amounts of 
water formed within a C-type shock (Wardle et al.
1998; Lockett et al. 1998).  OH masers adjacent to compact H\textsc{ii} regions 
are produced by photodissociation of water (Elitzur \& de Jong 1978; 
Hartquist \& Sternberg 1991; Hartquist et al.  1995), but in that case 
there is a strong dissociating FUV flux from the central star.  This 
dissociating flux is largely absorbed and reradiated by 
grains, providing sufficient FIR photons to also pump the 1665/7 
MHz transitions (e.g.  Pavlakis \& Kylafis 1996b).  This cannot be the 
case for the unaccompanied 1720 MHz masers associated with 
SNR-molecular cloud interactions.  Although only a weak UV field is 
required to dissociate the water (Lockett et al. 1998), it is difficult 
to deliver UV photons to the postshock gas, as the attenuation of an 
external flux by intervening dust generates a FIR 
continuum that inverts the main line transitions.

This problem is circumvented if the irradiation of the molecular cloud 
by X-rays produced by the hot gas in the interior of the adjacent SNR 
(Wardle et al.  1998) is taken into account.  This indirectly leads to 
dissociation: energetic electrons are photoejected by the X-rays and 
thermalise by collisions with molecules.  Some of the collisions 
excite the Werner Ly-$ \alpha $ band of H$_2$, and the subsequent 
radiative decay contributes a photon to the local FUV radiation field.  
Because X-rays are much more penetrating than UV and the ratio of the 
grain to molecule absorption cross-sections for X-rays is much lower 
than for UV photons (e.g.  Maloney, Hollenbach \& Tielens 1996) very 
little FIR is generated by this mechanism.

\begin{figure}[!t]
		\plotone{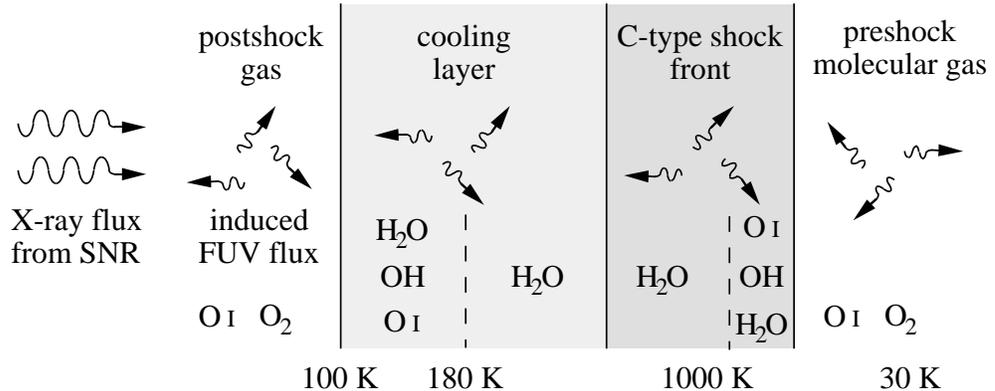}
		\caption{A model for the 
		production of OH in molecular clouds associated with supernova 
		remnants.  A C-type shock is driven into a molecular cloud 
		adjacent to a SNR. The X-ray flux from the SNR interior (to 
		the left) permeates the cloud, inducing a weak secondary FUV 
		flux that is produced locally throughout the cloud.  The shock 
		efficiently wave incorporates atomic and molecular oxygen into 
		water.  Once the shocked gas cools, water is subsequently 
		dissociated to OH and then to O\textsc{i} by the secondary UV flux.
	\label{fig:model}}
\end{figure}

A sketch of the model is shown in Fig. \ref{fig:model},
which shows a C-shock driven into a molecular cloud.  The preshock 
gas at the right of Fig.  \ref{fig:model} is at rest, and the oxygen 
that is not bound in CO exists as O\textsc{i} or O$_2$.  As the gas is 
accelerated, compressed, and heated within the shock front the atomic 
and molecular oxygen is rapidly converted to water.  The shocked, 
water-rich gas cools as it drifts behind the shock front.  The entire 
shock structure is subject to the weak dissociating UV flux produced 
by the X-rays permeating the cloud.  The dissociations cannot compete 
with the rapid reformation of water until the temperature drops to 
below about 200 K, at which point water is dissociated to OH and 
subsequently to O\textsc{i}.

\section{H$_2$\ $ 2 \micron $ emission associated with Sgr A East}

The masers to the southeast of Sgr A East, at positions B and D--F in 
Fig.  \ref{fig:YZ96masers}, arise in shocked gas in the M-0.02-0.07 
cloud, where the morphology and velocity gradients have already been 
interpreted in terms of an interaction with Sgr A East (Serabyn et al. 1992).  
The maser at position C on the northwest boundary of Sgr A East is
likely to be indicative of a similar interaction.  At position A the 
circumnuclear ring appears to be disturbed by gas associated with the 
Northern Arm (Jackson et al. 1993; Yusef-Zadeh et al. 1996).

In the case of Sgr A East there is sufficient information to think 
about the interaction in more detail.  The hot interior of Sgr A East, 
which supplies the overpressure driving the shock wave into the 
M-0.02-0.07 cloud, should be in rough pressure equilibrium with the 
magnetic pressure inferred from Zeeman measurements, $4\times 10^{-7} 
\mathrm{\,erg \, cm}^{-3}$, which dominates the pressure in the 
shocked molecular gas.  Indeed, ASCA observations of the X-ray 
emitting gas filling Sgr A East (Koyama et al.  1996) imply $ n_e 
\approx 6 \mathrm{\,cm}^{-3} $ and $ T \approx 10 \mathrm{\,keV} $, or 
a pressure of $ 2\times 10^{-7} \mathrm{erg\,cm}^{-3} $.  Further, the 
X-ray luminosity from Sgr A East, $ 10^{36} \mathrm{\,erg\,s^{-1}} $, 
is sufficient to produce OH behind the shock front.  Equating the 
magnetic pressure, $ 4\times 10^{-7} \mathrm{\,erg\, cm}^{-3} $, to $ 
\rho v_s^2 $, where $ \rho $ is the preshock density and $ v_s $ is 
the shock speed, and adopting a preshock density of $ 
n_{\mathrm{H}}=2\times 10^4 \mathrm{\,cm^{-3}} $ (Mezger et al.  
1989), we infer $ v_s \approx 25$--$30 \mathrm{\,km\,s^{-1}} $.  A 
C-type shock at this speed produces an intensity in the 1-0 S(1) line 
of $ 10^{-4} $--$10^{-3} \mathrm{\,erg\,s^{-1}\,cm^{-2}} $ (Draine et 
al.  1983; Kaufman \& Neufeld 1996).

\begin{figure}[!t]
	\plotone{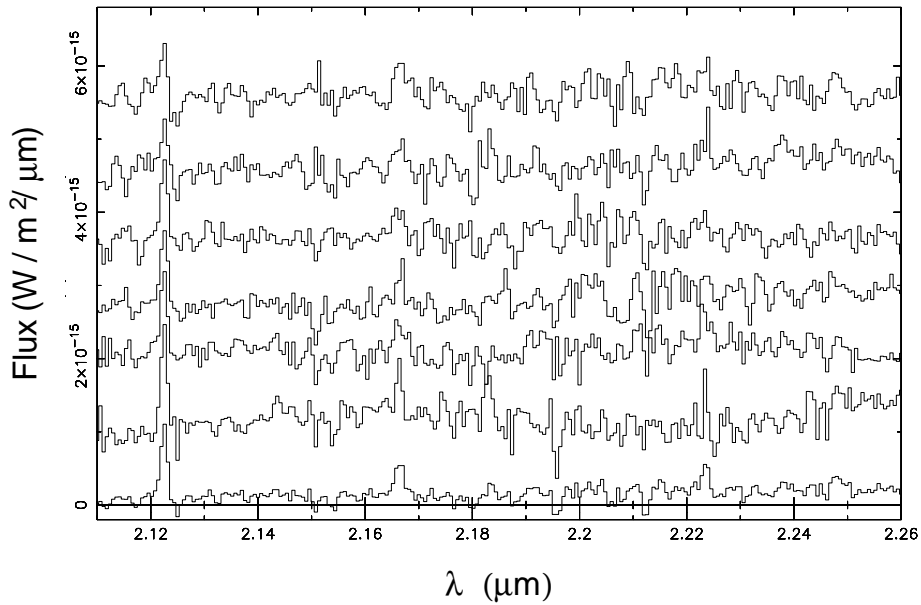} 
	\caption{Six adjacent spectra, with vertical 
	scales offset from one another, extracted from a CGS4 observation 
	of the OH(1720 MHz) maser spots in Sgr A East and (at the bottom) 
	the averaged spectrum.  Each spectrum is of an area $ 
	1.2'' \times 2.5'' $.  The second row from the top coincides with 
	position A, $\alpha(1950) = 17^{h} 42^{m} 33.6^{s}, 
	\delta(1950) = -29^{0} 00' 09.6''$.  
	The top row is 1.2~arcsec SW, and the 3rd - 6th spectra are 1.2, 2.5, 
	3.7, and 4.9 arcsec to the NE of the above coordinates.}
	\label{fig:h2}
\end{figure}

This should be observable.  Recently we used the long slit 1-5$\mu$m 
spectrometer, CGS4, at UKIRT to search for shocked H$_2$ emission 
associated with the OH(1720 MHz) masers in the 50 km/s cloud 
(covering positions A and D--G in Fig.  \ref{fig:YZ96masers}).  The 
spectra we obtained are shown in Fig.  \ref{fig:h2}.
The 1-0 S(1) (2.122~$\mu$m) line can be seen in the individual 
spectra, but other lines of H$_{2}$ are too weak for them to be 
securely identified.  The 1-0 S(0) (2.223~$\mu$m) and 2-1 S(1) 
(2.248~$\mu$m) lines of H$_{2}$, and H I Br~$\gamma$ (2.166~$\mu$m) 
are, however, evident in the averaged spectrum.

The intensity of the 1-0 S(1) line, assuming that $ A_K \approx 2.5 $, 
is $ 1.5\times 10^{-4} \mathrm{\,erg\,s^{-1}\,cm^{-2}} $, consistent 
with the levels ``predicted'' above.  However, the the 2-1 S(1) / 1-0 
S(1) and 1-0 S(0) / 1-0 S(1) line ratios in the average spectrum are 
about 1/7 and 1/3 respectively, somewhere between those expected for 
collisional excitation and for pure fluorescence.  These ratios 
could be produced by intense UV irradiation of molecular gas 
(Sternberg \& Dalgarno 1989), but the UV radiation field associated 
with Sgr A East is likely to be much less than in the inner parsec of the 
Galaxy, and the maser spots are distributed in an area free of the 
H\textsc{ii} regions that would be associated with massive star 
formation.  Given the weak detections of the 2-1 S(1) and 1-0 S(0) 
lines, further observations are required to determine the line ratios 
accurately.

Preliminary results from AAT and HST observations of shocked H$_2$ 
associated with the masers are discussed elsewhere in these 
proceedings (see the contribution by Yusef-Zadeh et al.).

\section{Discussion}

In principle a low-energy cosmic-ray flux enhanced by a factor of 100 
over the solar neighbourhood value has a similar effect on the heating 
and chemistry of molecular gas as X-ray irradiation from adjacent SNR 
(e.g.  Maloney et al.  1996).  In this case, energetic electrons 
produced by cosmic ray ionisation play an analogous role to the 
electrons photoejected by X-rays.  This could well apply to many 
galactic SNRs -- for example, the high-energy cosmic-ray flux in the 
W44, W28 and IC 443 SNRs that is required to produce the associated EGRET 
sources is roughly two orders of magnitude larger than the solar 
neighbourhood value (Esposito et al.  1996), so a naive extrapolation 
down to the cosmic-ray energies important for ionisation suggests that 
the cosmic-ray and X-ray contributions to the secondary FUV flux may 
be similar.  In the case of Sgr A East, Melia et al. (1998) have argued 
that the Galactic Centre EGRET source 2EG J1746-2852 can be explained 
similarly (see also the contribution to these proceedings by Fatuzzo 
et al.).

\acknowledgements
The CGS4 spectra were obtained as part of the UKIRT Service Programme.  
UKIRT is operated by the Joint Astronomy Centre on behalf of the U.K. 
Particle Physics and Astronomy Research Council.  The Special Research 
Centre for Theoretical Astrophysics is funded by the Australian 
Research Council under the Special Research Centres programme.  F. 
Yusef-Zadeh's work was supported in part by NASA grant NAGW-2518.


\begin{references}
\reference Backer, D. C. 1988, 
	in AIP Conf Proc 174: Radio Wave Scattering in the Interstellar Medium, 
	ed. J. Cordes, B. Rickett \& D. Backer (New York: AIP), 111
\reference Claussen, M. J., Frail, D. A., Goss, W. M. \& Gaume, R. A. 1997, 
	\apj, 498, 143
\reference Draine, B. T., Roberge, W. G. \& Dalgarno, A. 1983, 
	\apj, 264, 485
\reference Elitzur, M. 1976, 
	\apj, 203, 124
\reference Elitzur, M. 1998, 
	\apj, 504, 390
\reference Elitzur, M. \& de Jong, T. 1978,
	\astap, 67, 323
\reference Elitzur, M., Hollenbach, D. J. \& McKee, C. F. 1989,
	\apj, 346, 983
\reference Esposito, J. A., Hunter, S. D., Kanbach, G. \& Sreekumar, P. 1996,
	\apj, 461, 820
\reference Frail, D. A., Diamond, P. J., Cordes, J. M. 
	\& van Langevelde, H. J. 1994,
	\apjl, 427, L43
\reference Frail, D.A., Goss, M.W. \& Slysh, V.I. 1994, 
	\apjlett, 424, L111
\reference G\"usten, R., Genzel, R., Wright, M. C. H., Jaffe, D. T., 
	Stutzki, J. \& Harris, A. I. 1987, 
	\apj, 318, 124
\reference Hartquist, T. W., Menten, K. M., Lepp, S. \& Dalgarno, A. 1995,
	\mnras, 272, 184
\reference Hartquist, T. W. \& Sternberg, A. 1991,
	\mnras, 248, 48
\reference Hollenbach, D. J. \& McKee, C. F. 1989, 
	\apj, 342, 306
\reference Jackson, J.  M., Geis, N., Genzel, R., Harris, A.  I., Madden,S., 
	Poglitsch, A., Stacey, G. J. \& Townes, C. H. 1993, 
	\apj, 402, 173
\reference Kaufman, M. J. \& Neufeld, D. A. 1996, 
	\apj, 456, 611
\reference Killeen, N. E. B., Lo, K. Y. \& Crutcher, R. 1992, 
	\apj, 385, 585
\reference Koralesky, B., Frail, D. A., Goss, W. M., Claussen, M. J. 
	\& Green, A. J. 1998, 
	\aj, in press
\reference Koyama, K., Maeda, Y., Sonobe, T., Takeshima, T., Tanaka, Y. 
	\& Yamauchi, S. 1996, PASJ, 48, 249
\reference Lazio, T. J. W. \& Cordes, J. M. 1998, 
	\apj, in press (astro-ph/9804157)
\reference Lockett, P., Gauthier, E. \& Elitzur, M. 1998,
	\apj, in press
\reference Maloney, P. R., Hollenbach, D. J. \& Tielens, A. G. G. M. 1996, 
	\apj, 466, 561
\reference Melia, F., Fatuzzo, M., Yusef-Zadeh, F. \& Markoff, S. 1998, 
	\apjl, 508, L65ÐL69
\reference Mezger, P. G., Zylka, R., Salter, C. J., Wink, J. E., Chini, R., 
	Kreysa, E. \& Tuffs, R. 1989, 
	\astap, 209, 337
\reference Pavlakis, K. G. \& Kylafis, N. D. 1996a, 
	\apj, 467, 300
\reference Pavlakis, K. G. \& Kylafis, N. D. 1996b, 
	\apj, 467, 309
\reference Plante, R. L., Lo, K. Y. \& Crutcher, R. M. 1995, 
	\apjlett, 445, L113
\reference Schwarz, U. J., Lasenby, J., Kronberg, P. P. \& Wielebinski, R. 1990, 
	in Galactic and Intergalactic Magnetic Fields, 
	ed. R. Beck (Dordrecht : Kluwer ), 383
\reference Serabyn, E., G\"usten, R., Walmsley, C. M., Wink, J. E. 
	\& Zylka, R. 1986, 
	\astap, 169, 85
\reference Serabyn, E., Lacy, J. H. \& Achtermann, J. M. 1992, 
	\apj, 395, 166
\reference Sternberg, A. \& Dalgarno, A. 1989, 
	\apj, 338, 197
\reference van Langevelde, H. J., Frail, D. A., Cordes, J. M.
	\& Diamond, P. J. 1992,
	\apj, 396, 686
\reference Wardle, M., Yusef-Zadeh, F. \& Geballe, T. R. 1998
	\apjl, submitted (astro-ph/9804146)
\reference Yamauchi, S., Kawada, M., Koyama, K., Kunieda, H.
	\& Tawara, Y. 1990, 
	\apj, 365, 532
\reference Yusef-Zadeh, F., Cotton, W., Wardle, M., Melia, F.
	\& Roberts, D. A. 1994, 
	\apjl, 434, L33
\reference Yusef-Zadeh, F., Roberts, D. A., Goss, W. M., Frail, D. A. 
	\& Green, A. J. 1998,\apj, in press (astro-ph/9809279)
\reference Yusef-Zadeh, F., Roberts, D.A., Goss, W.M., Frail, D. \&
	Green, A. 1996, 
	\apj, 466, L25

\end{references}
\end{document}